\documentclass{appolb}
\usepackage{epsfig,amsmath,amsfonts,amssymb,cite}
\usepackage{bm}% bold math
\newcommand{\be}{\begin{equation}}
\newcommand{\ee}{\end{equation}}
\newcommand{\ba}{\begin{eqnarray}}
\newcommand{\ea}{\end{eqnarray}}

\begin{document}
\title{Selected issues on justification of holographic approach to QCD%
\thanks{Paper contribution to ExcitedQCD2010.}%
}
\author{S. S. Afonin
\address{V. A. Fock Department of Theoretical Physics, Saint-Petersburg State University,
1 ul. Ulyanovskaya, 198504, Russia.}
}
\maketitle
\begin{abstract}
Some problems with theoretical foundations of bottom-up holographic
models are briefly discussed. It is pointed out that the
spectroscopic aspects of these models in principle do not require the AdS/CFT
prescriptions and may be interpreted as just an alternative language
expressing the phenomenology of QCD sum rules in the large-$N_c$
limit. A general recipe for incorporation of the chiral symmetry
breaking scale into the soft-wall holographic models is proposed.
\end{abstract}

\section{Introduction and some general discussions}

At the present time the bottom-up holographic models for QCD (called
also AdS/QCD models) have become very popular approach to
description of phenomenology of strong interactions. An interesting
question appears if it is possible to understand the 5-dimensional
models in terms of traditional theoretical methods, {\it i.e.}
without appealing to the AdS/CFT correspondence or any other ideas
from the string theory? The main purpose of the given report is to
demonstrate that at least in one important case the answer seems to
be positive. The so-called soft-wall models introduced in\cite{son}
reproduce correctly the phenomenology of QCD sum rules in the
large-$N_c$ limit. One can try to invert the underlying logic of the
soft-wall models: start from a Lagrangian describing an infinite
number of free stable mesons with fixed quantum numbers which are
expected in the large-$N_c$ limit of QCD, fix the Regge form of
spectrum without corrections\footnote{Although such corrections are
expected on the theoretical grounds\cite{we} we simplify the matter
in order to arrive at holographic-like models in the shortest way.},
require to reproduce the analytic structure of the Operator Product
Expansion for the two-point correlation functions, and use the
Kaluza-Klein reduction, {\it i.e.} the fact that a free 5D field is
equivalent to the infinite number of free 4D fields whose masses are
determined by 5D background. It turns out that one arrives then
uniquely at the soft-wall models, other possibilities are less
satisfactory in the phenomenology\cite{afonin}. In addition, when
integrating back the fifth coordinate, a boundary term does not
vanish for the vector and (in one particular case) scalar fields and
the emerging term may be interpreted as a source for those fields
thus justifying the use of AdS/CFT conjecture for the given cases.
The soft-wall models in their spectroscopic applications may be
consequently viewed as just an alternative language expressing the
phenomenology of QCD sum rules in the large-$N_c$ limit with ensuing
accuracy and number of input parameters. The details of the
corresponding analysis are given in Ref.\cite{afonin}. In the
present short contribution, we will dwell on a couple of comments
which were not discussed in\cite{afonin}.

The observation above waives the following objection against the
holographic models\cite{cohen}: QCD has an infinite number of
operators with any set of quantum numbers and in general these
operators mix. One could therefore expect that an arbitrarily large
subset of them may contribute to any given process because there is
no obvious suppression scale. The approximation of lowest-dimension
operators that is typically used for constructing holographic models
looks thus {\it ad hoc}. We note that this argument is equally
applicable to the standard QCD sum rules and for some reasons the
approximation above works usually very well. As long as the
bottom-up holographic models are deeply related to the QCD sum
rules, it is not surprising that the same restriction to a minimal
set of the simplest operators probing the quantum numbers of
interest is successful in the holographic approach.

Unfortunately, the relationship of bottom-up holographic models with
the planar sum rules entails the following property: There is no
internal criterium indicating whether we may trust the results or
not, the goodness of a holographic model is always checked {\it a
posteriori} by comparison with the phenomenology. The original QCD
sum rules possess such a criterium --- the existence of a "Borel
window" (a region of stability of Borel parameter), but this
criterium is lost when taking the limit of large-$N_c$.

The next comment concerns the foundations of AdS/QCD models. The
whole approach is often criticized for its speculative nature. Which
specific assumptions have no solid ground and which are reasonable?
An optimistic point of view that is widespread among practitioners
of holographic models for QCD consists in a belief that the AdS/QCD
correspondence is a reasonable approximation to holographic duals of
QCD because the latter is approximately conformal in the UV limit.
Thus the conformal symmetry of AdS metric ensures the right behavior
of correlators and what is left to speculations is how to set the
confinement scale and how to model chiral symmetry breaking. We find
this belief poorly justified. First of all, QCD at large space-like
momenta is weakly coupled, this feature is opposite to what is
required by the conjecture of AdS/CFT correspondence which, taking
at face value, would lead to a strongly coupled holographic dual in
the UV limit, hence, the semiclassical approximation fails for the
latter. Moreover the large-$N_c$ limits of QCD and of field theory
in Maldacena's example\cite{mald} are quite different ($\alpha_sN_c$
is not constant in the latter but grows with $N_c$), let alone the
fact that the operators in QCD have anomalous dimensions\footnote{We
note incidentally that if the bottom-up models are regarded as a 5D
reformulation of QCD sum rules then this problem does not look
dangerous.} contrary to the $N=4$ supersymmetric Yang-Mills theory
considered in\cite{mald}. This makes really questionable why the
AdS/QCD prescriptions may be used for approximate QCD duals without
modifications and even why we may hope at all that such duals exist.
Such a belief represents thus the main speculative assumption in the
whole enterprise.

One usually hopes, however, that if we are lucky enough in
appropriate breaking conformality in the IR region then we may
obtain nice 5D effective models for QCD using the AdS/CFT
correspondence as a guiding principle. In other words, we may try to
use the conformal symmetry of QCD at high energies as the first
approximation and account for breaking of this symmetry by means of
some successful parametrization despite the fact that the breaking
of conformal symmetry shapes drastically the physics of strong
interactions. If this hope achievable? We are inclined to think that
rather no than yes. The matter is that within QCD there is an
explicit counterexample to such a programme. QCD possesses an
approximate chiral symmetry in the sector of light flavors which is
also broken. Nevertheless it is well known that the use of the
manifest chiral symmetry as the first approximation for building
low-energy effective models is not satisfactory whatever
parametrization accounting for the breaking of chiral symmetry is
applied. The reason is known as well --- the chiral symmetry of QCD
is realized in the Nambu-Goldstone mode at low energies and this
very realization must be used for model building as the first
approximation. The chiral symmetry breaking effect shapes
drastically the physics of strong interactions at low energies and
therefore by no means may be regarded as some correction to another
approximation. As to the highly excited states, the linear chiral
symmetry, most likely, is not restored (see, {\it e.g.}, discussions
in reviews\cite{noCS}). The lesson with the chiral symmetry seems to
suggest that a search for "another realization" of conformal
symmetry in hadron physics would be promising, the modern AdS/QCD
models are not in this line.

Although objections concerning the foundations of AdS/QCD models are
quite serious, the holographic methods are certainly interesting as
a language that allows to discuss within a uniform framework various
approaches to modeling the interactions and spectral characteristics
of light hadrons, heavy-light systems, hadron form factors, QCD
phase diagram, and other phenomenological aspects which were
previously the subjects of investigations for different communities.
This unifying property is remarkable indeed. As was shown in
Ref.\cite{afonin}, at least in some important applications one can
in principle dispense with the prescriptions dictated by the AdS/CFT
correspondence in constructing 5-dimensional effective models for
QCD with the same final results. This is, however, possible at the
cost of alternative assumptions, the use of AdS/CFT prescriptions as
a guide turns out to be more compact and more elegant. In addition,
the holographic approach proposes an interesting new way for
description of the chiral symmetry breaking. We discuss briefly this
subject in the next section.

\section{Chiral symmetry breaking}

The incorporation of the Chiral Symmetry Breaking (CSB) into
5-di\-men\-sio\-nal effective hadron models has a lot of freedom for
speculations because there are no quarks, hence no microscopic model
of CSB related directly to QCD may be constructed, the best we may
do is to describe the consequences of CSB on the hadron level. The
first such consequence is the phenomenological fact that the masses
of parity partners which seemingly belong to the same chiral
multiplet are quite different. To reflect this effect one should
introduce a mechanism for this mass splitting. Within the 5D
framework, the chiral symmetry does not exist at all because there
is no analogue for the matrix $\gamma_5$ in five dimensions, the CSB
can be therefore simulated only indirectly by means of somewhat
different description of states with equal spin but opposite parity.

The second consequence of CSB consists in the appearance of massless
(in the chiral limit) pseudoscalar mesons due to the Goldstone
theorem. The both consequences should be described with the help of
a mechanism that explains, {\it e.g.}, why the relation $m_{\pi}=0$ is
naturally related with $m_{a_1}^2\gtrsim 2m_{\rho}^2$ (instead of
$m_{a_1}=m_{\rho}$ as naively expected from the linear chiral
symmetry). The simplest such a mechanism that is realized in the
AdS/QCD models is borrowed from
the low-energy effective field theories: One introduces a scalar
field $X$ that acquires a non-zero vacuum expectation value (v.e.v.)
$X_0(z)$ and is coupled to the axial-vector field $A_{M}$ through
the covariant derivative, $D_M X=\partial_M X-ig_5A_MX$. The minimal
part of 5D action describing CSB is
\be
\label{82}
S_{\text{CSB}}=\int d^4 x\int dz
\sqrt{|G_{MN}|}e^{\Phi}\left(|D_M X|^2-m_X^2|X|^2\right).
\ee
It is sufficient to retain the quadratic in fields part
because the equations of motion can provide a
non-zero v.e.v. already in this case if the bulk space is curved.
The term~\eqref{82} must be added to a basic action of a holographic
model in question, for instance such an action for the soft-wall model
may be written in the form~\cite{afonin}
\be
\label{7}
S_{\text{5D}}=(-1)^{J}\int d^4x\,dz e^{\Phi}
a^{-2J+3}
\left\{(\partial_{\mu}\varphi_J)^2-(\partial_z\varphi_J)^2-
m_J^2a^2\varphi_J^2\right\},
\ee
where the gauge $\varphi_{z\dots}=0$ is accepted for the gauge fields
$\varphi_J$ of arbitrary spin $J$, the function $a(z)$ parametrizes the
metric, $ds^2=a^2(z)(dx_{\mu}dx^{\mu}-dz^2)$, and the function $\Phi(z)$
represents a dilaton background. The equation of motion for action~\eqref{7}
yielding the mass spectrum can be cast into the form of
Schr\"{o}dinger equation,
\be
\label{25}
-\psi_n''+U\psi_n=m_n^2\psi_n,
\ee
\be
\label{25b}
U=\frac{\Phi''}{2}+\left(\frac{\Phi'}{2}\right)^2+
\left(\frac32-J\right)\dfrac{\Phi'a'+a''+\left(\frac12-J\right)\frac{(a')^2}{a}}{a}+a^2m_J^2,
\ee
where $\varphi_n=e^{-\Phi/2}a^{J-3/2}\psi_n$ and the prime denotes the derivative with respect to $z$.

The equation of motion determining the v.e.v. $X_0$ is
nothing but the equation~\eqref{25} for a massless scalar
particle. According to a recipe based on the AdS/CFT
correspondence~\cite{kleb}, it must behave at $z=0$ as
$X_0(z)\vert_{z\rightarrow0}=C_1z+C_2z^3$
where $C_1$
is associated with the current quark mass, $C_1\sim m_q$, and $C_2$
with the quark condensate, $C_2\sim\langle\bar{q}q\rangle$. This
interpretation implies a somehow well established correspondence of
the model to QCD. As we do not have such a rigorous correspondence,
it would be more honest to say that the incorporation of the 4D massless
scalar particle can be related to the spontaneous appearance of two order
parameters with mass dimension one and three and this property may
be exploited to mimic the CSB.

In fact, there is the third consequence of the CSB in QCD: The
effective emergence of the scale $\Lambda_{\text{CSB}}\simeq4\pi
f_{\pi}\approx1\div1.2$~GeV. The physics of strong interactions is
known to be substantially different below $\Lambda_{\text{CSB}}$ and
above $\Lambda_{\text{CSB}}$. This consequence of the CSB is not
incorporated into the existing AdS/QCD models. In what follows we
propose a possible scheme for incorporation of
$\Lambda_{\text{CSB}}$ into the holographic models~\cite{afonin}.

The equation of motion for $X$ represents a second order linear
differential equation that has two independent solutions,
\be
\label{86}
X_0(z) = C_1X_1(z)+C_2X_2(z).
\ee
In general $X_0(z)$ is not normalizable, namely
the solution $X_1(z)$ spoils the
normalizability at $z\rightarrow0$ while (in the soft wall models)
$X_2(z)$ does at $z\rightarrow\infty$. We can require the normalizability
for $X_0(z)$ and construct the following normalizable solution,
\be
\label{87} X_0(z) = C_2X_2(z)\vert_{z\leq
z_0}+C_1X_1(z)\vert_{z>z_0}.
\ee
As long as the fifth coordinate $z$ is associated with the inverse energy
scale in the holographic approach, it looks natural to interpret $z_0^{-1}$
as the scale $\Lambda_{\text{CSB}}$ because the physics below and above
this scale will be different.

In the case of the soft-wall model with positive-sign dilaton background
the corresponding solutions are
\be
\label{89} X_1(z)=ze^{-z^2}U(-1/2,0;z^2),\qquad X_2(z)=z^3e^{-z^2}M(1/2,2;z^2),
\ee
where $U$ and $M$ stay for the $U$ and $M$ Kummer functions.
The "potential"~\eqref{25b} has the following asymptotics for the axial-vector
mesons,
\be
\label{95}
U\vert_{z\gg
z_0}=z^2\left(1+g_5^2C_1^2e^{-2z^2}\right)+\frac{3}{4z^2},
\,\,\,
U\vert_{z\ll
z_0}=z^2\left(1+g_5^2C_2^2z^2\right)+\frac{3}{4z^2},
\ee
while for the vector case $C_1=C_2=0$. This
behavior demonstrates that the spectrum of axial-vector states
approaches rapidly to the spectrum of vector mesons.
The rate of such a "Chiral Symmetry Restoration"
is exponential that is in agreement with the analysis~\cite{we}
based on the QCD sum rules. This qualitative feature differs from
the typical predictions within the soft-wall
models where the shift between the vector and axial-vector
masses square tends to a constant.

\section*{Acknowledgments}

The work is supported by the Alexander von Humboldt Foundation and
by RFBR, grant 09-02-00073-a.

\end{document}